\documentclass[conference]{IEEEtran}
\IEEEoverridecommandlockouts
\usepackage{amsmath,amssymb,amsfonts}
\usepackage{algorithmic}
\usepackage{graphicx}
\usepackage{textcomp}
\usepackage{xcolor}
\usepackage{booktabs}
\usepackage{url}
\usepackage{comment}
\usepackage{cleveref}
\usepackage[numbers]{natbib}
\usepackage{tablefootnote}
\usepackage{dcolumn,tabularx,ragged2e}

\usepackage{defs}

\begin{document}

\title{A customizable approach to assess software quality through Multi-Criteria Decision Making}

\author{\IEEEauthorblockN{Francesco Basciani}
\IEEEauthorblockA{\textit{University of L'Aquila}\\
L'Aquila, Italy \\
francesco.basciani@univaq.it}
\and
\IEEEauthorblockN{Daniele Di Pompeo}
\IEEEauthorblockA{\textit{University of L'Aquila}\\
L'Aquila, Italy \\
daniele.dipompeo@univaq.it}
\and
\IEEEauthorblockN{Juri Di Rocco}
\IEEEauthorblockA{\textit{University of L'Aquila}\\
L'Aquila, Italy \\
juri.dirocco@univaq.it}
\and
\IEEEauthorblockN{Alfonso Pierantonio}
\IEEEauthorblockA{\textit{University of L'Aquila}\\
L'Aquila, Italy \\
alfonso.pierantonio@univaq.it}
}

\maketitle

\begin{abstract}
Over the years, Software Quality Engineering has increased interest, demonstrated by significant research papers published in this area.
Determining when a software artifact is qualitatively valid is tricky, given the impossibility of providing an objective definition valid for any perspective, context, or stakeholder.
Many quality model solutions have been proposed that reference specific quality attributes in this context.
However, these approaches do not consider the context in which the artifacts will operate and the stakeholder's perspective who evaluate its validity.
Furthermore, these solutions suffer from the limitations of being artifact-specific and not extensible.

In this paper, we provide a generic and extensible mechanism that makes it possible to aggregate and prioritize quality attributes.
The user, taking into account his perspective and the context in which the software artifact will operate, is guided in defining all the criteria for his quality model.
The management of these criteria is then facilitated through Multi-Criteria Decision Making (MCDM).
In addition, we present the PRETTEF model, a concrete instance of the proposed approach for assessing and selecting MVC frameworks.

\end{abstract}

\section{Introduction}\label{sec:intro}

In modern software products, the definition of quality is a critical and complex process due to several properties that can be considered. 
Furthermore, the selection and prioritization of these properties to measure are strictly related to the context to which the software refers.
In addition, stakeholders can decide on preferences for these properties based on their feelings and expertise.

Over the years, many studies have been conducted to provide valuable tools to define the quality of software products~\cite{Bertoa:2010td,lopez2014assessing,DBLP:journals/jss/CortellessaPET22,ma2013assessing,SEAA2021}.
Still, this multitude of studies demonstrates how much the debate on establishing software quality is a hot topic.
Many of these studies offer reasonable solutions for some stakeholders or specific contexts without considering the context in which the stakeholders find themselves can change and their point of view, effectively making these approaches useless due to their specificity.

This paper aims to provide a generic and extensible mechanism by which assertions about the quality of software artifacts can be made.
It extends~\cite{quatic-bash, basciani2019tool} by providing a tool to define a quality model for Multi-Criteria Decision Making (MCDM) and managing criteria involved in the quality assessment of software products. 
Furthermore, the expressiveness provided by the MCDM approach allows for establishing priorities in a set of established criteria.

Indeed, introducing the concept of MCDM in a generic and extensible system, such as the Quality Metamodel~\cite{basciani2019tool}, provides a highly customizable tool that aims to determine the quality of a software product according to stakeholders' point of view.

We validate our framework on Model-View-Control (MVC) framework selection by introducing the MCDM quality model extension.
It is worth to remark technology investment is crucial to keep IT teams updated with current trends. 
Therefore, choosing the right software technologies depends on various quantitative and qualitative factors.
Indeed, our framework helps IT teams select the right MVC framework by defining and extracting quality metrics.
PREsentation, Trend, TEchnology, and Features (PRETTEF) quality attributes have been combined to
select a suitable MVC framework that reflects the users’ needs.

In this respect, the contributions of this work are as follows: 
\begin{itemize}
\item an extension of the tool-supported approach defined in~\cite{quatic-bash, basciani2019tool} to support (MCDM)~\cite{umm-e-habiba_survey_2009,aruldoss_survey_2013} assessments; 
\item the PRETTEF model, a concrete instance of the extended approach for selecting the suitable MVC web developing framework. 
\end{itemize}
The structure of the paper is as follows: \Cref{sec:background} presents some backgrounds for our work. The proposed approach is illustrated in \Cref{sec:methodology}. 
The next section explains the evaluation settings used to study the approach’s performance. 
\Cref{sec:usecase} presents the experimental results of applying our approach to assess the best suitable web developing framework.
\Cref{sec:related} reviews the related work and associates them with our approach. 
Finally, \Cref{sec:conclusion} outlines possible future work and draws some conclusions.
 \section{Background}\label{sec:background}
This section describes how we dealt whit the quality assessment of a software artifact in our previous work~\cite{basciani2019tool}.
Also, it provides a brief introduction to MCDM approaches for selecting the most suitable alternative whit respect to a set of criteria by highlighting the most relevant aspects.

\subsection{Quality models}

Assessing the quality of software systems is underpinned by the definition of several metrics that, once evaluated and properly aggregated, can give insights into the quality of the analyzed system.
One way to formulate a software quality model is to identify a small set of high-level attributes and decompose them into sets of subordinate attributes \cite{quatic-bash}.
That approach is generic and extensible, with a tool-chain enabling users to specify custom quality models whose quality attributes are hierarchically aggregated. Their evaluation depends on metrics specifically conceived and applied to the artifacts to be analyzed and combined. 

\begin{figure*}[!htb]
	\centering
	\includegraphics[width=1\linewidth]{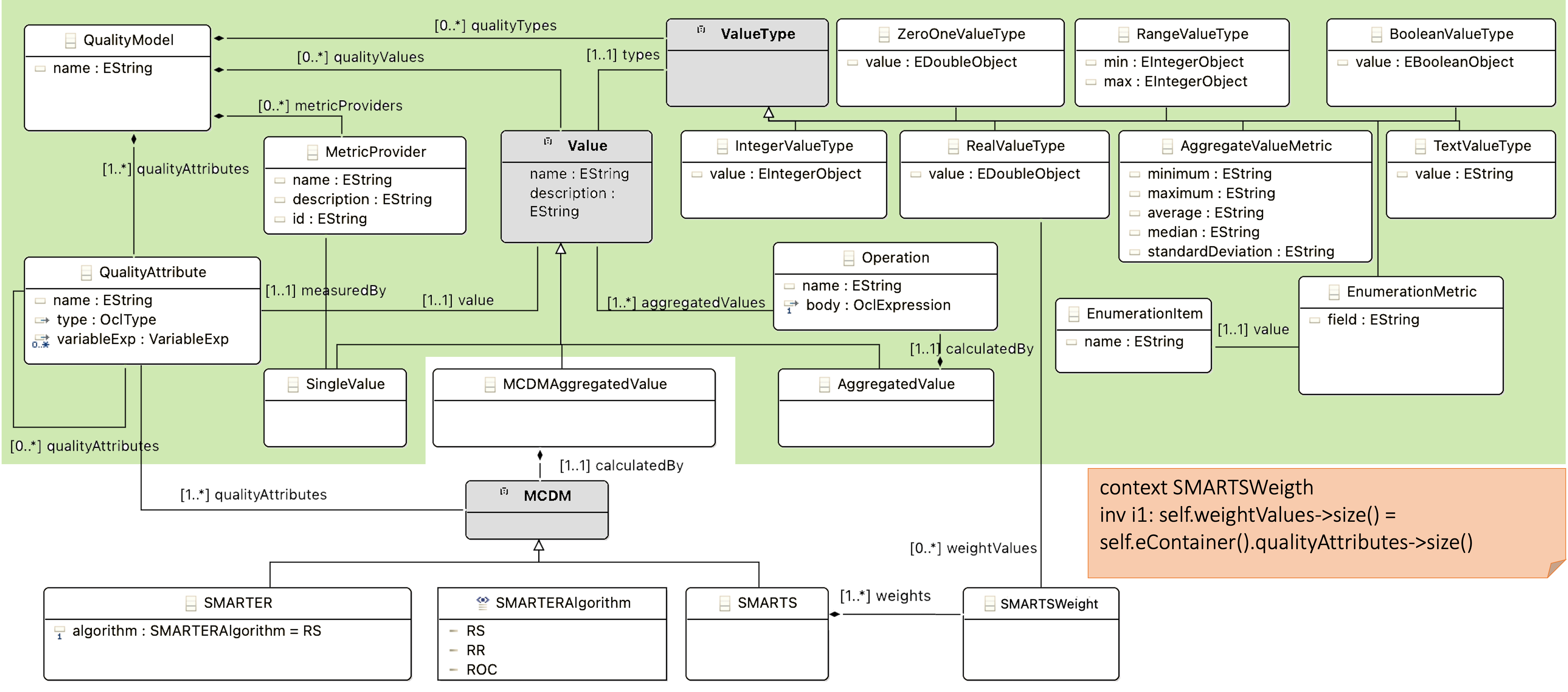}
	\caption{The MCDM Quality Metamodel}
\label{fig:QMDiagram}
\end{figure*}

For this purpose, using a Quality Metamodel plays a key role since it enables the specification of quality models according to user requirements.
We based our work on enhancing the approach proposed in \cite{quatic-bash,basciani2019tool} and Fig.~\ref{fig:QMDiagram} shows how we extended the Quality Metamodel. 
Specifically, the elements already present in the starting Quality Metamodel are in the green area while all other elements represent the extension provided.

As can be seen in the figure, the \texttt{QualityModel} represents the root element, which consists of \texttt{QualityAttribute}s, \texttt{Value}s,\texttt{ValueType}s, and \texttt{MetricProvider}s.
A \texttt{QualityAttribute} represents a quality aspect considered relevant for contributing to the quality assessment of a given artifact and it can be an aggregation of other quality attributes.
Thus, each quality attribute specifies how to aggregate the contained attributes to provide overall quality value for the considered attribute.
Each quality attribute has a Value representing its calculated Value that is abstract, and it can be \texttt{SingleValue} or \texttt{AggregatedValue}.
The metaclass \texttt{SingleValue} represents the Value obtained by applying a given \texttt{MetricProvider}, which refers to the software component's ability to calculate specific metrics.
\texttt{AggregatedValue} indicates a composition of different values specified through an OCL\footnote{\url{https://projects.eclipse.org/projects/modeling.mdt.ocl}} expression contained in the attribute body of the Operation element.
Each Value has a reference with a \texttt{ValueType} element, which is abstract and defines its type. Several specializations are provided to enable the specifications of integer, boolean or ranged values (e.g., from 0 to 5) for example (the other specializations have been omitted but can be viewed in \cite{quatic-bash,basciani2019tool}).

\subsection{MCDM}

In different application domains, Multi-Criteria Decision-Making (MCDM)~\cite{umm-e-habiba_survey_2009,aruldoss_survey_2013} approaches are used to make decisions when heterogeneous factors are involved in the evaluation of many alternatives. Just to mention a few, economics~\cite{zavadskas2011multiple,kou2014evaluation}, medical~\cite{salabun2017comparative,ren2017thermodynamic}, and non-functional analysis~\cite{kumar2017review, kabak2014prioritization,SEAA2022} domains took benefits by applying MCDM approaches. 

The purpose of an MCDM is to assist the decision-maker in mastering the different involved criteria and advancing toward selecting a possible alternative. Among the factor that impacts the selection of suitable alternatives, the definition of the priorities/weights on a set of criteria is a key aspect of the MCDM approaches that strongly rely on subjective evaluation. 
In a generic decision-making process, the following entities can be identified: 
\begin{itemize}
    \item \emph{Objective}: it defines the purpose for which the artifact assessment is carried out; therefore, the result will have to get as close as possible to the set goal.
    \item \emph{Decision-maker (\DM)}: he is the actor who has the task of defining the selection criteria and the method of evaluating alternatives. In most cases, it is not a single person but a group of people;
    \item \emph{Alternatives}: the alternatives represent the space of solutions;
    \item \emph{Criteria}: they represent the attributes on which the assessment is performed. The criteria describe measurable characteristics;
    \item \emph{Assessment}: it computes the ranked list of the alternatives consistent with \DM value judgments.
    \item \emph{Solution}: it represents the alternative that achieves the best compromise in relation to the objective.\end{itemize}
Over the years, several approaches~\cite{aruldoss2013survey} have been developed to deal with MCDM assessments, \eg the SMARTER and SMARTS ones~\cite{triantaphyllou_multi-criteria_2000,velasquez_analysis_2013}. 
Moreover, many hybrid approaches~\cite{hu2014exploring,wu2012ranking,tzeng2007evaluating} have been proposed to combine two or more existing ones.
The goal of hybrid MCDM approaches is to use different MCDM techniques to combine the evaluation of different criteria.
In the remainder of the section, we focus on SMARTER and SMARTS since they underpin the approach proposed in the next section where such approaches are combined with the quality model proposed in the previous section.

\subsubsection{SMARTER} \label{sec:smarter}
The Simple Multi-Attribute RaTing Exploiting Ranks (SMARTER)~\cite{edwards1994smarts} technique consists of a combined approach between the subjective elicitation of preferences with an objective distribution of weights.
This approach relies on the following rationale: ``\textit{The \DM usually can rank items much more easily than give weight to them}''.
Therefore, the elicitation phase does not provide for the assignment of weights by the decision maker. In this approach, the decision maker's task is limited to ranking the criteria according to her preferences. In contrast, the weight of the individual criteria are assigned through predefined methods~\cite{Roszkowska13}, i.e., Rank Order Centroid (ROC), Rank Sum (RS), Rank Reciprocal (RR). These methods help the \DM develop more consistent and transparent weights according to their rank assignments. This method takes those ranks as inputs and converts them to weights for each item. 

Given a list $n$ of prioritized ranked criteria, where each criterion $j$ has a rank $r_j$, $j \in (1,2,...,n)$
For each such method, a criterion weight assessment $w_j$ is defined in the following:

\begin{equation}
    w_j (ROC) = \frac{1}{n}\sum\limits_{k=j}^n \frac{1}{r_k}, j=1,2,...n
\end{equation}

\begin{equation}
    w_j (RR) = \frac{\frac{1}{r_j}}{\sum\limits_{k=1}^n \frac{1}{r_k}}, j=1,2,...n
\end{equation}

\begin{equation}
    w_j (RS) = \frac {n-r_j+1} {\sum\limits_{k=1}^n n-r_k + 1} = \frac {2(n+1-r_j)}{n(n+1)}, j=1,2,...n
\end{equation}

Once the weights have been assigned by one of the previous techniques, the utilities of alternatives $A = \{a_1, a_1, ..., a_m\}$ are evaluated by the following formula:
\begin{equation}
    u(a_i) = \sum \limits_{j=1}^n w_j * v_j(a_i)\end{equation}
where $n$ are the number of criteria, $w_j$ is the assigned weight for the criterion $j$, and $v_j(a_i)$ is the values of criteria $j$ in the alternative $i$.

\subsubsection{SMARTS}\label{sec:smarts}
The Simple Multi-Attribute Rating Swing Weight (SMARTS)~\cite{edwards1994smarts} method drives the \DM in selecting an alternative. Unlike the SMARTER approach, the \DM directly assigns ratings to the criteria on a natural scale. Then, SMARTS uses swing-weight techniques to select the most appropriate alternative. In particular, given a $W = \{w_1,w_2,...,w_n\}$  \DM ratings of the criterion, the following formula normalizes those values:
\begin{equation}
    w_{j(norm)} = \frac{w_j}{\sum\limits_{k=1}^n w_k}
\end{equation}
Holding the weighting of criteria and rating of alternatives as separate as
possible, the different scales of criteria are converted to a
common internal scale. Then, for each alternative $a_i$ the utilities $u(a_i)$ is computed as following:
\begin{equation}
    u(a_i) = \sum \limits_{j=1}^n w_{j(norm)} * sv_j(a_i)
\end{equation}
where $sv_j(a_i)$ is the scaled value of the criteria $j$ in the alternative $i$.

\begin{comment}
\subsubsection{AHP} 
The Analytic Hierarchy Process (AHP) is a robust and flexible MCDM methodology, which formulates the decision problem in a hierarchical structure and defines the priorities of its elements, each level, comparing their mutual importance for a common attribute. Many works present entangled approaches that 
\end{comment}
 \section{Proposed approach}\label{sec:methodology}

This section presents a process for solving MCDM decision problems using an extension of the quality model defined in~\cite{basciani2019tool,quatic-bash}.
In the MDCM approaches, the \DM performs subjective (e.g., the selection of criteria and their weight) and objective choices (i.e., the assessment of the alternatives) to achieve predefined goals. 
MCDM can be seen as a tool for evaluating alternatives when several criteria are considered to produce a common output. 
In previous studies~\cite{basciani2019tool,quatic-bash}, we presented a generic approach aiming at the custom definition of quality models where hierarchically organized quality attributes, whose evaluation depends on measurable aspects and applied on the software artifacts to be analyzed, are therefore combined.

Both domains (\ie quality models and MCDM) strongly rely on subjective choices that the user underpins to perform qualitative assessments and alternative evaluations. On the one hand, the quality models allow the modeler to define their choices by a generic, extensible, and self-contained language. On the other hand, the MCDM approach supports the \DM in assessing alternatives by well-defined and formal steps.

\subsection{Extended quality model notation}
We extended the quality model notation and tools to native support MCDM reasoning. 
First, we investigate possible mappings between the concepts belonging to the two domains. The quality model lets the user define single/aggregated attributes and the \DM aggregates criteria. For instance, hybrid MCDM approaches aggregate  criterion by involving different MCDM approaches to perform the alternative assessments, and the quality model aggregates quality attributes during the model actualization. 
As defined in \Cref{sec:background}, the quality values are aggregated by user-defined operations, and the quality model doesn't provide any support in the definition of the aggregation formula.

The \DM defines the objectives and the list of alternatives, whereas the quality model defines the criteria and the assessment settings. Then, the quality model evaluation engine assesses the alternatives.
Given an alternative and an input quality model, the automated assessment step applies the MCDM approaches to produce the score of the alternative. Such a score is used to rank the alternatives. Finally, the ranked list of alternatives is provided to the \DM.

In this context, as we introduced in Sec.~\ref{sec:background}, MCDM approaches have been proposed (\eg SMARTER, SMARTS, etc.) to support the \DM with predefined formulas in the alternative assessment. 

Given that the Quality Model allows the user to define hierarchical structures of quality attributes and how single measurement values have to be combined to achieve overall quality statements, we aim to extend the quality model to support hybrid MCDM approaches, where different criteria can be aggregated at different levels. 

\begin{figure}
	\centering
	\includegraphics[width=1\linewidth]{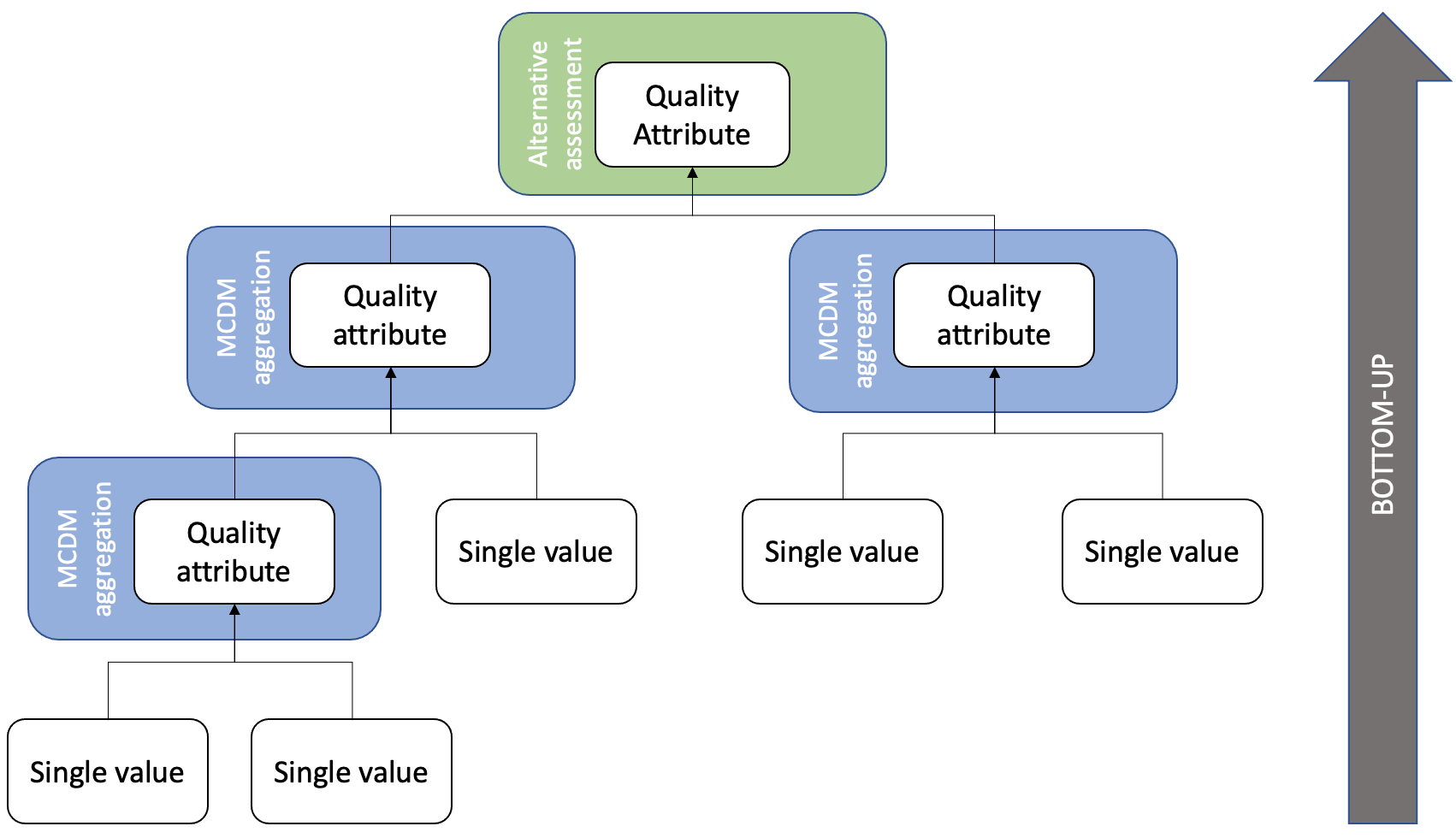}
	\caption{Combining MCDM approaches with a hierarchical quality model assessment.}
	\label{fig:hyerarchicalMCDM}
\end{figure}

The quality evaluation engine computes and aggregates the criteria in a bottom-up manner. Figure~\ref{fig:hyerarchicalMCDM} shows how different aggregating criteria can be expressed with the quality model. In particular, the alternatives assessment takes a bottom-up approach, starting from the elements at the bottom of the hierarchy, where a leaf represents a measurable value. The hierarchy is climbed up by aggregating the values with MCDM techniques. This process ends when the root is reached. The root of the hierarchy stores and represents the score assigned to an alternative.

To ensure that aggregate values could be defined through multiple criteria, as you can see in Fig. \ref{fig:QMDiagram} outside the gree area, the Quality Metamodel was enhanced by introducing the \texttt{MCDMAggregateValue} metaclass as an extension of \texttt{Value}.

This metaclass allows the introduction of the calculation of the \texttt{MCDM}.
Unlike the \texttt{AggregatedValue}, it calculates the aggregation of different \texttt{Value}s through MCDM techniques.

We, therefore, introduced SMARTS and SMARTER metaclasses as extensions to the abstract \texttt{MCDM} metaclass:
\begin{itemize}
    \item in the \texttt{SMARTER} metaclass, it is possible to indicate the type of algorithm to be used in the calculation through an enumerator, \texttt{SMARTERAlgorithm}), with the values: \texttt{RS}, \texttt{RR}, and \texttt{ROC}. (see Sec. \ref{sec:smarter}). 
    The chosen algorithm will be applied to the \texttt{QualityAttributes} to which this metaclass refers.
    \item the algorithm used for \texttt{SMARTS} is unique and is always \texttt{SWING} (see Sec. \ref{sec:smarts}). As anticipated in Section~\ref{sec:smarts} the SMARTER approach requires the \DM directly assigns ratings to the criteria in a natural scale. For this reason, an instance of \texttt{SMARTSWeight} allows the \DM define the ratings for each quality attribute involved in the assessment. It is worth noting that the \DM ratings should map the involved quality attributes. For this reason, the metamodel invariant (defined as an OCL statement in Figure~\ref{fig:QMDiagram}) ensures that the size of the score list is equal to the size of quality attributes.
\end{itemize}

Both approaches are represented by a metaclass (nominally SMARTER and SMARTS) calculating the aggregate value, each through a specific process.
More precisely, SMARTER's use requires indicating the algorithm type used for the aggregation (\texttt{SMARTERAlgorithm}): RS, RR, or ROC.
As a result, it returns the QualityAttributes aggregation as an ordered list.
With SMARTS, on the contrary, there is no need to indicate the algorithm to be used as this is implicit (SWING). Still, it is also necessary to provide at least one initializer, represented by the metaclass SMARTInitializer to return the final value given by the aggregation of at least one QualityAttribute.

\subsection{Extending the Quality model engine to support MCDM analysis}

The support engine for multi-criteria analysis has been implemented by providing built-in formulas defined in OCL. Therefore, this engine can interpret these formulas without further modifications.
In other words, the system takes as input the quality model defined by the user, considers the possible implementations of MCDM (i.e., SMART and SMARTER), and instantiates them on the artifacts object of evaluation.
This is possible thanks to the fact that the quality model engine supports the computation of aggregated values written using OCL.
Consequently, it was possible to reuse the existing calculation engine by focusing on the representation of SMARTER or SMART formulas via OCL.

Figure \ref{fig:hyerarchicalMCDM} shows the hierarchical organization of a model using the MCDM techniques. 
The engine, in a bottom-up manner, therefore, starting from the Single Values represented by the leaves of the tree structure, aggregates all the values up to providing a series of alternatives from which the decision maker can then determine the solution.

 \section{Automated MCDM reasoning for selecting web development frameworks}\label{sec:usecase}
This section presents an application of our approach in the selection process of an MVC framework. 
In particular, we instantiate the quality model on a concrete example of a user/company evaluating which MVC framework deserves more attention.
PREsentation, Trend, TEchnology, and Features (PRETTEF) quality attributes have been combined to select a suitable MVC framework that reflects the users' needs. 

\begin{figure}[t]
	\centering
	\includegraphics[width=1\linewidth]{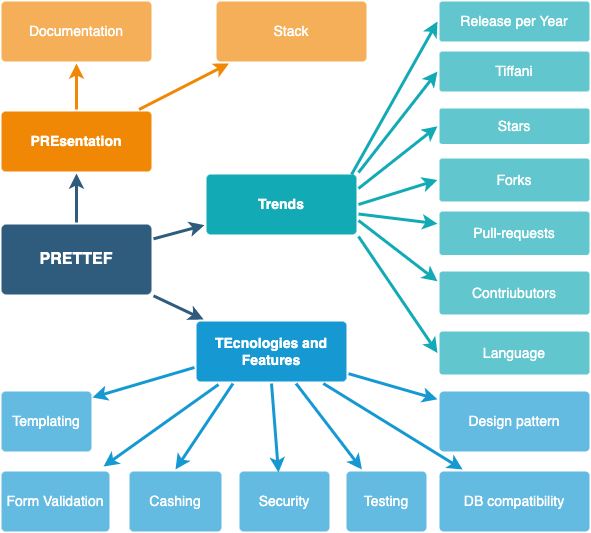}
	\caption{The PRETTEF model}
	\label{fig:prettef}
\end{figure}

\Cref{tab:alternatives} lists the 17 selected MVC frameworks that play as alternatives in the MCDM analysis. 
Those frameworks are written in 7 programming languages (\ie Java, PHP, JavaScript, Python, Ruby, and C\#), where JavaScript is the most used language. All the considered frameworks have more than 3,000 forks and more than 900 pull requests.
\Cref{fig:prettef} shows how the quality attributes and metrics are hierarchically organized in the PRETTEF model to evaluate the framework described above. 
Each node is a quality attribute with a corresponding value type in the model. The PRETTEF model uses predefined value types provided by the quality model, \eg numerical, categorical, and boolean values.
For instance, Stack and Design Pattern are defined as categorical values, Stars and Forks as numerical data, and Documentation as a boolean value.
To enable the assessment, the user/modeler has to assign numerical scores to non-numeric values (e.g., categorical and boolean). For instance, the current PRETTEF model uses  0.5, 0.5, and 1 for the back end, front end, and full-stack, respectively. 

Once the quality attributes are defined with their type, we apply SMARTER and SMARTS MCDM techniques for each aggregation layer by instantiating the corresponding \texttt{SMARTER} and \texttt{SMARTS} metaclasses. Both instances include the list of quality attributes involved in the assessment. SMARTER includes the weight specification algorithm, while SMARTS the list of weights. For instance, Table~\ref{tab:trends} lists the adopted rank and scores for SMARTER and SMARTS approaches, respectively.
It is worth noting that we defined our weights and ranks in the PRETTEF models. All the parameters, as well as the aggregation items, can be changed according to the modeler's needs. 
Once the modelers complete the specification of the enhanced quality, each alternative is evaluated among all the configurations.
\begin{table*}[]
    \centering
    \caption{Selected alternatives. Data are updated to Dec 2020}
    \label{tab:alternatives}
    \footnotesize
    \begin{tabular}{|l|l|r|l|r|r|r|}
    \hline
         Framework & URL & language & Licence & \#Forks & \#Pull-Requests \\ \hline
         
Spring MVC & \url{https://spring.io/projects/spring-framework} 	& Java 			& Apache 2.0 	&	20,550			&	3,393 \\ \hline
Laravel & \url{https://laravel.com}  	& PHP 			& MIT 			& 	16,685			&	3,433 \\ \hline
Symfony & \url{https://symfony.com}	& PHP 			& MIT 			&	7,105 			&	20,470 \\ \hline
Cake PHP & \url{https://cakephp.org}	& PHP   		& MIT 			&	3,406 			&	8,134 \\ \hline
Code Igniter& \url{https://codeigniter.com} & PHP 			& MIT 			&	7,731 			&	2777 \\ \hline
AngularJs & \url{https://angularjs.org} 	& JavaScript	& MIT			&	28,888			&	7,881 \\ \hline
React.JS & \url{https://reactjs.org}	& JavaScript	& MIT			&   25,206			&	8,459 \\ \hline
Express.js & \url{https://expressjs.com} 	& JavaScript	& MIT 			&	7,576 			& 	904 \\ \hline
Node.js  & \url{https://nodejs.org}	& JavaScript	& ISC 			&	14,420 			&	18,559 \\ \hline
Ember.Js  & \url{https://emberjs.com}  & JavaScript	& MIT			&	4,169 			&	8,631 \\ \hline
Backbone.js & \url{https://backbonejs.org} & JavaScript 	& MIT 			&	5,687 			&	1,831 \\ \hline
Vue.js 	& \url{https://vuejs.org}	& JavaScript	& MIT 			&	21,547 			&	1,597 \\ \hline
Django 	& \url{https://www.djangoproject.com}	& Python 		& Django o.s.	& 	18,870			&	11,713 \\ \hline
Flask 	& \url{https://flask.palletsprojects.com}	& Python		& BSD-3C		&	12,850			&	1,565 \\ \hline
Rails 	& \url{https://rubyonrails.org}	& Ruby 			&MIT 			&	17,751			&	24,042 \\ \hline
AspNetCore & \url{https://learn.microsoft.com/en-us/aspnet/core/}	& C\# 			& Apache 2.0  	& 	3,186			&	3,476 \\ \hline
Bootstrap & \url{https://getbootstrap.com}	& Javascript	& MIT			&	66,668			&	10,068\\ \hline

    \end{tabular}
\end{table*}

\begin{table}[]
    \centering
    \caption{Rank and weights for Trend quality attribute aggregation}
    \label{tab:trends}
    \footnotesize
    \begin{tabular}{|c|l|l|} \hline
QualityAttribute & Rank & Weight \\ \hline
GitHub Contributors & 3 & 0,111  \\ \hline
GitHub Stars & 3  & 0,111  \\ \hline
GitHub Pull Request & 2 & 0,166  \\ \hline
Forks & 1 & 0,335  \\ \hline
Release per Year & 3  & 0,111  \\ \hline
Language & 2 & 0,167 \\ \hline
    \end{tabular}
    
\end{table}
    
\begin{figure}
	\centering
	\includegraphics[width=\linewidth]{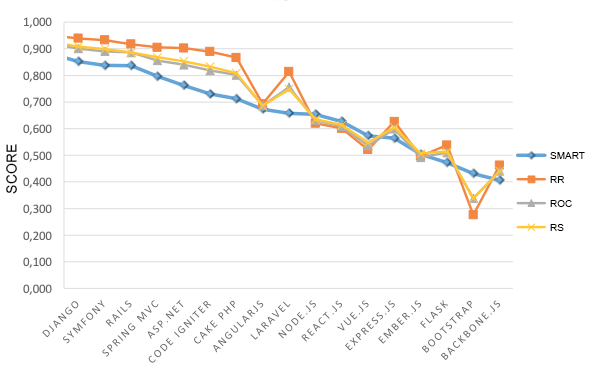}
	\caption{The alternative assessment}
	\label{fig:ev1}
\end{figure}

\begin{figure}
	\centering
	\includegraphics[width=\linewidth]{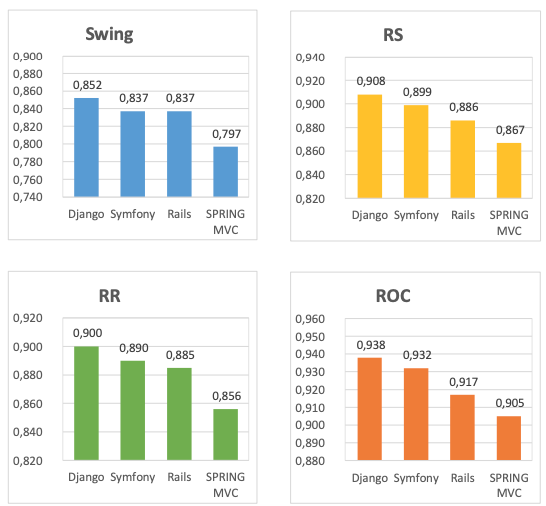}
	\caption{Top 4 alternatives}
	\label{fig:ev2}
\end{figure}

Figure~\ref{fig:ev1} shows how the RR, ROC, and RS algorithms have very similar trends: there are differences in evaluation only for the last items in the alternative rank. The SMART results are more subjective, and there is a more progressive trend among the alternatives  (that depends on the DM weights). It can also be seen that the result of the first 8 alternatives coincides for all four methods. At the same time, there is a different alternative rank between the swing technique and the RR on the AngularJs and Laravel frameworks. In the last 6 alternatives, ROC, RS, and RR trends fluctuate and correspond to different alternatives assessments.
Figure~\ref{fig:ev2} shows that the evaluation of the solutions is equivalent even using different techniques for calculating the weights. The ranking of the first five alternatives is comparable, even using various techniques. This evidence allows us to state that the methods based on the ordering are similar and can be used according to the decision maker's preference. At the same time, the Swing Weights technique can radically change the evaluation result. Therefore it is advisable to use the SMARTS method in the presence of experts in the field of application of the investigation context.

It is worth noting that this section aims at providing a  \section{Related work}\label{sec:related}

Baroni et al.\footnote{member of the QUASAR research team \url{https://sites.google.com/site/quasarresearchgroup/}} introduced the Meta-Model Driven Measurement (M2DM) that allows the specification of specific object-oriented design metrics through OCL constructs~\citep{baroni02}. 
The M2MD can be applied to formalize the MOOD2 metrics over the GOODLY design language metamodel~\cite{baroni02}, define object-oriented design metrics~\cite{baroni03}, express relational database schema metrics~\cite{baroni04}, and assess components composition using the CORBA Components Metamodel~\cite{GA05}. 

Abandah and Alsmadi showed that software metrics could be used in deficiency prediction~\cite{abandah2013call}.
For example, software call graph models were extracted and used to evaluate their ability to predict quality-related attributes in developed software products.
Specifically, the generated call graph models represent and reflect the degree of complexity, especially in understandability, testability, and product maintenance efforts. 
The extracted metrics are investigated in relation to or in correlation with bugs collected from reports.
They are used as input to a data miner for classification, prediction, and association analysis.

The ISO/IEC 9126 standard defines a quality model consisting of \emph{factors}, \emph{criteria}, and \emph{metrics}.
A factor (e.g., reliability or functional capacity) is decomposed into \emph{criteria} (e.g., safety or modularity).
Criteria are decomposed in terms of \emph{metrics} computed on the source code. 

The SQUALE project~\cite{mordalmanet:inria-00637364} extended the ISO/IEC 9126 standard by the \emph{practice} concept, which is an intermediate level between criteria and metrics.
This additional concept is exploited to assess if the considered project respects technical principles, e.g., the fact that complex classes should be more documented than trivial ones. 
Furthermore, the \emph{practices} concept defines new practices (and eventually composes them) to assess the quality of a software system by directly analyzing the source code through a supporting tool.
Finally, the SQUALE project introduced more than 50 practices collected in a list that is not intended to be closed. 

Siavvas et al.~\cite{siavvas2017qatch} presented the QATCH tool-chain to generate software quality models that reflect user specification through static analysis.
Furthermore, QATCH generates quality models that exploit a weighted average aggregation scheme to calculate the overall quality.
Expert judgments can define these weights.

A quality model specifically conceived to measure the quality of metamodels is introduced in~\cite{Bertoa:2010td}.
It allows for defining quality attributes, e.g., maintainability, portability, and usability, along with sub-characteristics, e.g., analyzability, adaptability, and understandability. 

EMF Metrics~\cite{Arendt:2010wb} supports the specification and calculation of metrics for models developed atop the Eclipse Modeling Platform (EMF). 
It focuses on metrics explicitly defined to measure the model's structural characteristics.
For example, it supports the number of attributes, classes, and associations.

In this paper, we introduced a quality computational framework that helps users compute their quality metrics on sever artifacts. 
Furthermore, we differ from the above methods because we introduced an extensible framework, which is also agnostic to the used modeling language. \section{Conclusion}\label{sec:conclusion}

In this work, we have defined a quality model (i.e., PRETTEF) integrated with Multiple Criteria Decision Making (MCDM) for the assessment of selecting Web Development frameworks based on software qualities. 
Furthermore, PRETTEF allows a decision-maker to define and modify criteria and metrics to make the analysis flexible. 
In fact, a decision maker can modify: i) the Attributes, Metrics, and Weights calculation, ii) the elicitation techniques, and iii) the data aggregation method.

Our approach eases the decision-making process and regulates the decision-makers subjectivity.
The proposed quality model combines classic MCDM methods, e.g., SMARTER and SMARTS techniques. 
The quality model effectively formalizes and structures the evaluation criteria or attributes.

 \section*{Acknologement}
The authors would like to thank Andrea Serafini for his hard work on supporting the PRETTEF use case.
\bibliographystyle{IEEEtran}
\bibliography{bib}

\end{document}